\documentclass{article}
\usepackage{psfig}
\begin{document}
\title{\vspace{-3cm}
\LARGE\bf Nonlinear QM as a fractal Brownian motion with
complex diffusion constant}
\author{Carlos Castro$^1$, Jorge Mahecha$^2$ and Boris
Rodr\'{\i}guez$^2$\\
{\small\em $^1$Center for Theoretical Studies of Physical
Systems,}\\ {\small\em Clark Atlanta University, Atlanta,
Georgia, USA}\\
\smallskip
{\small\em $^2$Institute of Physics, University of
Antioquia, Medell\'{\i}n, Colombia}}
\date{\today}

\maketitle

\begin{abstract}

A new nonlinear Schr\"odinger equation is obtained
explicitly from the fractal Brownian motion of a massive
particle with a complex-valued diffusion constant.
Real-valued energy (momentum) plane wave and soliton
solutions are found in the free particle case. The
hydro-dynamical model analog yields another (new) nonlinear
QM wave equation with physically meaningful soliton
solutions. One remarkable feature of this nonlinear
Schr\"odinger equation based on a fractal Brownian
motion model, over all the other nonlinear QM models,
is that the quantum-mechanical energy functional
coincides with the field theory one.

\end{abstract}

\section{\bf Introduction}
\label{sec:intro}

The theoretical study of quantum chaos has been developed
mainly in two areas: The phenomenological characterization
of the spacing of the energy levels of bound and
quasi-bound quantum physical systems, whose main analytical
tool is the random matrix theory \cite{mehta}, and the
semi-classical limit of chaotic classical systems
\cite{gutzwiller}. The semi-classical approach pretends to
seek solutions of the Schr\"odinger equation and to read in
the wave functions any fingerprints of classical chaos. Due
to the linearity of the Schr\"odinger equation there is no
place where the sensibility to the initial conditions can
be made manifest, which is present in nonlinear chaotic
systems. The Riemann zeta function has been considered as a
unifying link between those two approaches \cite{keating}.

We believe that quantum chaos is truly a new paradigm in
physics associated with non-unitary and nonlinear QM
processes based on non-Hermitian operators (implementing
time symmetry breaking). This chaotic behavior stems
directly from the nonlinear Schr\"odinger equation without
any reference to the nonlinear behavior of the classical
limit. See \cite{perez}. For this reason, the genuine
quantum chaos should be exhibited only by systems whose
behavior is correctly described by a nonlinear
Schr\"odinger equation.

The nonlinear QM has a practical importance in different
fields, like condensed matter, quantum optics and atomic
and molecular physics; even quantum gravity may involve
nonlinear QM. Another important example is in the modern
field of quantum computing. If quantum states exhibit small
nonlinearities during their temporal evolution, then
quantum computers can be used to solve NP-complete (non
polynomial) and \#P problems in polynomial time. Abrams and
Lloyd \cite{abrams} proposed logical gates based on non
linear Schr\"odinger equations and suggested that a further
step in quantum computing consists in finding physical
systems whose evolution is amenable to be described by a
NLSE.

On other hand, we consider that Nottale and Ord's
formulation of quantum mechanics \cite{nottale} from first
principles based on the combination of scale relativity and
fractal space-time is a very promising field of future
research. In this work we extend Nottale and Ord's ideas to
derive the nonlinear Schr\"odinger equation. This could
shed some light on the physical systems which could be
appropriately described by the nonlinear Schr\"odinger
equation derived in what follows.

The contents of this work are the following. In section
\ref{sec:hydro} we derive different nonlinear
Schr\"odinger-like equations starting from purely
hydro-dynamical considerations. In section
\ref{sec:nottale} a review of the derivation of the
Schr\"odinger equation, based on Nottale and Ord's
\cite{nottale} model of QM as a fractal Brownian motion of
a particle zigzagging back and forth in space-time, is
presented. In section \ref{sec:diffusion} we derive the
nonlinear Schr\"odinger equation from an extension of the
Nottale's approach to the case of a fractal Brownian
motion with a complex diffusion constant. In section
\ref{sec:soliton} real-valued energy solutions of the
nonlinear Schr\"odinger equation are proposed. In the final
section \ref{sec:conclu}, we summarize our conclusions and
include some additional comments.

\section{\bf Nonlinear Srh\"odinger equations based on
hydrodynamics}
\label{sec:hydro}

In this section we will write down two NLSE (nonlinear
Srh\"odinger equations) using the hydro-dynamical models of
QM \cite{bohm,madelung}. The first equation is based by
adding a hydrostatic pressure term to the Euler-Lagrange
equations \cite{pardy} and the second equation is obtained
by adding, instead, a kinematic pressure term. As far as we
know, this second equation has not appeared in the
literature before.

The hydrostatic pressure experienced by a fluid element at
a point $\vec{r}$ due to the force of gravity is given by
the Euler equation $-\vec{\nabla}p=\rho\vec{g}$;
$\vec{\nabla}$ is the ordinary gradient, $p$ the pressure,
$\rho$ the density and $\vec{g}$ the acceleration of
gravity. For example, if the density and acceleration are
uniform one can integrate such equation and arrived at
$p=\rho g x$ giving the pressure at a given depth $x$.

The author \cite{pardy} proposed to establish the QM analog
of the Euler equation by relating the density $\rho$ to the
quantum mechanical probability density $\psi^*\psi$ and by
integrating the equation. Setting $\rho=\psi^*\psi$, $b$ a
mass-energy parameter and the particular case that
$p=\rho$, then one has that the hydrostatic potential is
given by the integral
\begin{equation}
b\int\vec{g}(\vec{x})\cdot
d\vec{r} = -b\int\frac{\vec{\nabla}p}{\rho}\cdot d\vec{r} =
-b\ln\frac{\rho}{\rho_0} = -b\ln(\psi^*\psi),
\end{equation}
setting $\rho_0=1$. This is the nonlinear potential energy
induced from a hydrostatic pressure term.

It is important to normalize the logarithms by a constant
which we set to unity for convention. $-b\ln(\psi^*\psi)$
has energy units. From now these logarithmic terms are
normalized that way.

The hydrostatic pressure term \cite{pardy} has energy units
and explains in a straightforward fashion the nonlinear
term (nonlinear potential) added to the standard
Schr\"odinger equation by Bia{\l}ynicki-Birula and
Mycielski \cite{birula} long ago. The parameter $b$ has
units of mass (energy), so the nonlinear wave equation is
given by \cite{birula} after adding the nonlinear potential
$-b\ln(\psi^*\psi)$.

The Birula-Mycielski NLSE for a particle is
\begin{equation}
i\hbar\frac{\partial\psi}{\partial
t}=-\frac{\hbar^2}{2m}\nabla^2\psi + U\psi -
b[\ln(\psi^*\psi)]\psi.
\label{eq:birula}
\end{equation}
A derivation of this equation from the Nelson stochastic QM
was given by Lemos (\cite{elaf82} p. 615 and
\cite{lemos}).  As interesting as this equation may be
there are some problems. Such equation does not obey the
homogeneity condition (see Weinberg \cite{weinberg}) which
says that if $\vert\psi\rangle$ represents a physical
state, the rays $\vert\lambda\psi\rangle$ must also
represent the same physical state, for any complex constant
$\lambda$. But equation (\ref{eq:birula}) is not invariant
under $\psi$ $\to$ $\lambda\psi$ because the logarithmic
nonlinear potential breaks such homogeneity, this NLSE is
not scaled by $\lambda$.

Another problem is that plane wave solutions to equation
(\ref{eq:birula}) do not seem to have a physical
interpretation due to extraneous dispersion relations. Only
the soliton solutions were physically meaningful
\cite{pardy}. Upper limits on the values of the parameter
$b$ had been found to be \cite{kamesberg} $b<3\cdot
10^{-15}eV$, which correspond to an electron soliton width
of 3 mm \cite{pardy}. The smallness of $b$ is itself no
reason to disregard equation (\ref{eq:birula}) as
physically relevant. For the authors, another problem with
the logarithmic nonlinear potential term is that the
hydrostatic pressure term in the NLSE is given by an
explicit function of both $\psi$ and it complex conjugate
$\psi^*$. It is desirable to write a NLSE solely in terms
of the $\psi$ variable, or $\psi^*$, but not combined.
Thus, another new NLSE can immediately be written and/or
modified by adding the kinetic-pressure terms to the
Euler-Newton hydro-dynamical equations of motion, i.e. by
adding the term $(1/2)\rho V^2$ and taking
$\rho=a\psi^*\psi$; where $a$ is a mass parameter,
different from $b$, $\vec{V}=\vec{p}/m$ and $\vec{p}$ is
the momentum.

Using the relations from the Hamilton-Jacobi theory
\begin{equation}
\frac{\psi}{\psi^*}=e^{2iS(x)/\hbar},\quad
\vec{p}=\vec{\nabla}S(x)=m\vec{V},
\end{equation}
we can express the square of the velocity in terms of
$\psi$ and $\psi^*$ as follows,
\begin{equation}
\vec{V}=-i\frac{\hbar}{2m}\vec{\nabla}\ln\frac{\psi}{\psi^*},
\end{equation}
so the energy-density becomes
\begin{equation}
\frac{1}{2}\rho\vert\vec{V}\vert^2=\frac{a\hbar^2}{8m^2}\psi\psi^*
\vec{\nabla}\ln\frac{\psi}{\psi^*}\cdot\vec{\nabla}\ln\frac{\psi^*}{\psi},
\end{equation}
from which we immediately conclude that the corresponding
nonlinear potential term associated with the
kinematical pressure term is
\begin{equation}
\frac{a\hbar^2}{8m^2}
\vec{\nabla}\ln\frac{\psi}{\psi^*}\cdot\vec{\nabla}\ln\frac{\psi^*}{\psi}.
\end{equation}
Hence a candidate for a NLSE is
\begin{equation}
i\hbar\frac{\partial\psi}{\partial
t}=-\frac{\hbar^2}{2m}\nabla^2\psi + U\psi - b[\ln(\psi^*\psi)]\psi
+\frac{a\hbar^2}{8m^2}
\left(\vec{\nabla}\ln\frac{\psi}{\psi^*} \cdot
\vec{\nabla}\ln\frac{\psi^*}{\psi}\right)\psi.
\label{eq:newNLSE}
\end{equation}
Here the Hamiltonian is Hermitian and $a\ne b$ both are
mass-energy parameters to be determined experimentally. As
far as we know, this NLSE has not been derived so far.

The new term can be written also in the form
\begin{equation}
\vec{\nabla}\ln\frac{\psi}{\psi^*} \cdot
\vec{\nabla}\ln\frac{\psi^*}{\psi} =
-\left(\vec{\nabla}\ln\frac{\psi}{\psi^*}\right)^2.
\end{equation}

For simplicity purposes, from now on we shall refer to
these nonlinear potential kinematic pressure terms as those
terms of the type
\begin{equation}
U^{kin}\sim (\vec{\nabla}\ln\psi)^2.
\end{equation}
The reason we choose to impose that notation will become
clear in the next sections.

Our goal now is to derive NLSE directly from the fractal
space time dynamics of a particle undergoing a Brownian
random motion. And such fractal space time interpretation
does not require to add {\it ad-hoc\/} terms like

1. David Bohm's quantum potential \cite{bohm} into the
Hamilton equation. See also \cite{perez}.

2. To use the hydro-dynamical models discussed so far
\cite{pardy,bohm,madelung} nor to add the hydrostatic
pressure and kinematic pressure terms to the Euler-Newton
equations of motion as we have shown above, in an {\it
ad-hoc\/} fashion.

The new NLSE can be obtained from first principles if, an
only is, we assume a fractal trajectory of a point particle
associated with a Brownian random motion \cite{nottale}.

Before we begin, we deem it very important to add some more
comments about the kinematic pressure terms
\begin{equation}
\frac{1}{2}\rho V^2 \Leftrightarrow
\frac{\hbar^2}{2m}\frac{a}{m} \vert\vec{\nabla}\ln\psi\vert^2,
\end{equation}
versus the hydrostatic pressure terms
\begin{equation}
\int\frac{\vec{\nabla}p}{\rho} \Leftrightarrow -b\ln(\psi^*\psi)
\end{equation}
in the new NLSE.

The hydrostatic term $-b\ln(\psi^*\psi)\psi$ explicitly
breaks homogeneity $\psi\to\lambda\psi$ of the NLSE.
Whereas the kinematic pressure term $(\hbar^2)/(2m)(a/m)
\vert\vec{\nabla}\ln\psi\vert^2\psi$ does preserve the
homogeneity condition and the NLSE should scale with a
$\lambda$ factor, fact that can be easily verified.

The hydrostatic pressure term is not compatible with the
motion kinematics of a particle executing a fractal
Brownian motion. Only in the $m\to\infty$ limit (heavy
particle) it is reasonable to speak of the static limit.
See \cite{pardy} for a discussion of this limit. There are
another deeper reasons to derive the NLSE from an
underlying dynamics of a particle in a fractal space time,
or more simply, from a fractal trajectory in a fixed space
time background. We may, or may not, be switching on the
quantum gravitational aspects of space time, so deeply
linked to the non-linearity of QM. Perhaps the nonlinearity
of QM is deeply inter-wined with the quantum gravitational
aspects of a Cantorian-fractal-space-time
\cite{elnaschie,castro}.

Our goal is far less ambitious. Returning to our main
points, the two NLSE have introduced two additional
parameters of mass-dimension $a$, $b$ (or one parameter in
the special case $a=b$). Such parameters need to be found
experimentally. The advantage of the fractal formalism is
that we will be able to relate the $a$, $b$ parameters to
the Planck constant $\hbar$ itself, rather to have new
parameters in physics unrelated to $\hbar$.

\section{\bf QM as mechanics in non differentiable spaces}
\label{sec:nottale}

We will be following very closely Nottale's derivation of
the ordinary Scr\"odinger equation \cite{nottale}. The
readers familiar with this work may omit this section.
Recently Nottale and Celerier \cite{nottale} following
similar methods were able to derive the Dirac equation
using bi-quaternions and after breaking the parity symmetry
$dx^{\mu} \leftrightarrow - dx^{\mu}$, see references for
details. Also see the Ord's paper \cite{ord} and the
Adlers's book on quaternionic QM \cite{adler}. For
simplicity the one-particle case is investigated, but the
derivation can be extended to many-particle systems.

In this approach particles do not follow smooth
trajectories but fractal ones, that can be described by a
continuous but non-differentiable fractal function
$\vec{r}(t)$. The time variable is divided into
infinitesimal intervals $dt$ which can be taken as a given
scale of the resolution. If $\Phi(t,t',dt)$ is a smoothing
function centered on $t$, for example a step function of
width $2dt$, a continuous and differentiable approximation
to the true fractal $\vec{r}(t)$ can be constructed as
follows,
\begin{equation}
\vec{r}(t,dt)=\int\limits_{-\infty}^\infty\Phi(t,t',dt)\vec{r}(t')dt'.
\end{equation}
While $\vec{r}(t)$ = $\vec{r}(t,0)$ is non-differentiable,
any $\vec{r}(t,dt)$, called ``fractal trajectory'', is
differentiable for all $dt\ne0$.

Non-differentiability implies a lost of causality. For this
reason the fractal trajectories are good candidates to
describe the quantum behavior. Feynman in his path integral
formulation of QM already found a interesting result
involving the time scale $dt$: When seen at a time scale
$dt$, the quantum mechanical mean quadratic velocity of a
particle is $\langle v^2\rangle\propto 1/dt$. This can be
easily explained by an argument involving fractals. If the
trajectory is a fractal curve of fractal dimension $D$, the
space and time resolutions are related by $dt\propto dx^D$,
so that $\langle v^2\rangle\propto (dx/dt)^2\propto
dt^{2(1/D-1)}$. The comparison with Feynman's result leads
to $D=2$.

Here we omit the details of the arguments leading to
$dt\propto dx^D$, which the interested reader can find at
\cite{nottale} and references therein. They has as
ingredients the scale dependence of any fractal curve and
the renormalization group.

A fractal function $f(x,\epsilon)$ can have, besides the
derivative $\partial f(x,\epsilon)/\partial x$, the new
derivative with respect to the scale, $\partial
f(x,\epsilon)/\partial\epsilon$. It was found useful to use
$\ln\epsilon$ instead of $\epsilon$ as the variable for
resolution. Renormalization group arguments say that the
following relation is valid \cite{nottale},
\begin{equation}
\frac{\partial f(x,\epsilon)}{\partial\ln\epsilon} = a(x) +b
f(x,\epsilon),
\end{equation}
this means that the variation of $f$ under an infinitesimal
scale transformation $d\ln\epsilon$ depends only on $f$
itself. This differential equation can be integrated to
give us
\begin{equation}
f(x,\epsilon) = f_0(x)\left[ 1 +
\zeta(x)\left(\frac{\lambda}{\epsilon}\right)^{-b} \right].
\label{eq:fractfun}
\end{equation}
$\lambda^{-b}\zeta(x)$ is an integration constant and
$f_0(x) = -a(x)/b$. This result says that any fractal function
can be approximated by the sum of two terms, one
independent of the resolution and other resolution
dependent. Due to the resolution dependence is associated
to the fractal properties, and those are product of the
non-differentiability, then is expected that $\zeta(x)$ is a
fluctuating function with zero mean.

Provided than $a\ne0$ and $b<0$ two cases can be
considered: (i) $\epsilon\ll\lambda$, the scale dependent
term is dominant and $f$ is given by a scale-invariant
fractal-like power law with fractal dimension $D=b-1$,
namely $f(x,\epsilon)$ = $f_0(x)(\lambda/\epsilon)^{-b}$.
(ii) if $\epsilon\gg\lambda$ then $f$ becomes independent
of the scale. $\lambda$ is the de Broglie wave length.

A continuous but non-differentiable function $f(t)$ at $t$
has two possible values of the derivative at $t$, for this
reason its approximation by a fractal function requires
considering ``left'' and ``right'' derivatives. For the
position vector the following two infinitesimal differences
can be considered,
\begin{equation}
\begin{array}{rcl}
\vec{r}(t+dt,dt)-\vec{r}(t,dt) &=&\displaystyle \vec{b}_+(\vec{r},t)dt +
\vec{\xi}_+(t,dt)\left(\frac{dt}{\tau_0}\right)^\beta,\\
\vec{r}(t,dt)-\vec{r}(t-dt,dt) &=&\displaystyle \vec{b}_-(\vec{r},t)dt +
\vec{\xi}_-(t,dt)\left(\frac{dt}{\tau_0}\right)^\beta,
\label{eq:difference}
\end{array}
\end{equation}
where $\beta=1/D$, and $\vec{b}_+$ and $\vec{b}_-$ are
average forward and backward velocities \cite{nottale}.
Adopting the non standard analysis formulation, $dt$ is
also the time scale.

The instantaneous velocities are easily obtained from
equation (\ref{eq:difference}),
\begin{equation}
\vec{v}_\pm(\vec{r},t,dt) = \displaystyle \vec{b}_\pm(\vec{r},t) +
\vec{\xi}_\pm(t,dt)\left(\frac{dt}{\tau_0}\right)^{\beta-1},
\label{eq:velocities}
\end{equation}
In the quantum case, $D=2$, then $\beta=1/2$, so that
$dt^{\beta-1}$ is a divergent quantity, from which is
evident the non-differentiability.

Then, following the definitions given by Nelson in his
stochastic QM approach (Lemos in \cite{elaf82} p. 615; see
also \cite{lemos,ghirardi}), Nottale define mean backward
an forward derivatives as follows,
\begin{equation}
\frac{d_\pm\vec{r}(t)}{dt} = \lim\limits_{\Delta t\to\pm0}
\left\langle \frac{\vec{r}(t+\Delta t)-\vec{r}(t)}{\Delta
t} \right\rangle,
\end{equation}
from which the forward and backward mean velocities are
obtained,
\begin{equation}
\frac{d_\pm\vec{r}(t)}{dt} = \vec{b}_\pm.
\end{equation}

For his deduction of Schr\"odinger equation from this
fractal space-time classical mechanics, Nottale starts by
defining the complex-time derivative operator
\begin{equation}
\frac{\delta}{dt} = \frac{1}{2}\left(
\frac{d_+}{dt}+\frac{d_-}{dt}\right)
-i\frac{1}{2}\left(
\frac{d_+}{dt}-\frac{d_-}{dt}\right),
\end{equation}
which after some straightforward definitions and
transformations takes the following form,
\begin{equation}
\frac{\delta}{dt} = \frac{\partial}{\partial
t}+\vec{V}\cdot\vec{\nabla} - iD\nabla^2.
\label{eq:Notcomplder}
\end{equation}
$D$ is a real-valued diffusion constant to be related to
the Planck constant. Now we are changing the meaning of
$D$, since no longer a symbol for the fractal dimension is
needed, it will have the value $2$.

The $D$ comes from considering that the scale dependent
part of the velocity is a Gaussian stochastic variable with
zero mean, (see de la Pe\~na at \cite{elaf82} p. 428)
\begin{equation}
\langle d\xi_{\pm i}d\xi_{\pm j}\rangle = \pm 2 D \delta_{ij} dt.
\end{equation}
In other words, the fractal part of the velocity $\vec{\xi}$,
proportional to the $\vec{\zeta}$, amount to a Wiener
process when the fractal dimension is $2$.

Afterwards, Nottale defines a set of complex quantities
which are generalization of well known classical quantities
(Lagrange action, velocity, momentum, etc), in order to be
coherent with the introduction of the complex-time
derivative operator.

The complex time dependent wave function $\psi$ is
expressed in terms of a Lagrange action $S$ by
$\psi=e^{iS/(2mD)}$. $S$ is a complex-valued action but $D$
is real-valued. The velocity is related to the momentum,
which can be expressed as the gradient of $S$,
$\vec{p}=\vec{\nabla}S$. Then the following known relation
is found,
\begin{equation}
\vec{V}=-2iD\vec{\nabla}\ln\psi.
\end{equation}

The Schr\"odinger equation is obtained from the Newton's
equation (force = mass times acceleration) by using the
expression of $\vec{V}$ in terms of the wave function
$\psi$,
\begin{equation}
-\vec{\nabla}U = m\frac{\delta}{dt}\vec{V} =
-2imD\frac{\delta}{dt}\vec{\nabla}\ln\psi.
\end{equation}

Replacing the complex-time derivation (\ref{eq:Notcomplder})
in the Newton's equation gives us
\begin{equation}
-\vec{\nabla}U = -2im\left(D\frac{\partial}{\partial t}
\vec{\nabla}\ln\psi\right)
-2D\vec{\nabla}\left(D\frac{\nabla^2\psi}{\psi}\right).
\end{equation}
Simple identities involving the $\vec{\nabla}$ operator were
used by Nottale. Integrating this equation with respect to the
position variables finally yields
\begin{equation}
D^2\nabla^2\psi+iD\frac{\partial\psi}{\partial t}
-\frac{U}{2m}\psi = 0,
\end{equation}
up to an arbitrary phase factor which may set to zero. Now
replacing $D$ by $\hbar/(2m)$, we get the Schr\"odinger
equation,
\begin{equation}
i\hbar\frac{\partial\psi}{\partial t} +
\frac{\hbar^2}{2m}\nabla^2\psi = U\psi.
\end{equation}
The Hamiltonian operator is Hermitian, this equation is
linear and clearly is homogeneous of degree one under the
substitution $\psi\to\lambda\psi$.

\section{\bf Nonlinear QM as a fractal Brownian motion with
a complex diffusion constant}
\label{sec:diffusion}

Having reviewed Nottale's work \cite{nottale} we can
generalize it by relaxing the assumption that the diffusion
constant is real; we will be working with a complex-valued
diffusion constant; i.e. with a complex-valued $\hbar$.
This is our new contribution. The reader may be immediately
biased against such approach because the Hamiltonian ceases
to be Hermitian and the energy becomes complex-valued.
However this is not always the case. We will explicitly
find plane wave solutions and soliton solutions to the
nonlinear and non-Hermitian wave equations with real
energies and momenta.

For a detailed discussion on complex-valued spectral
representations in the formulation of quantum chaos and
time-symmetry breaking see \cite{petrosky}. Also a
complex-valued time and two-times (see \cite{bars} and
references therein) complex-valued dimensions have been
discussed in \cite{lapidus,elnaschie1}.

Nottale's derivation of the Schr\"odinger equation in the
previous section required a complex-valued action $S$
stemming from the complex-valued velocities due to the
breakdown of symmetry between the forwards and backwards
velocities in the fractal zigzagging. If the action $S$ was
complex then it is not farfetched to have a complex
diffusion constant and consequently a complex-valued
$\hbar$ (with same units as the complex-valued action).

Our derivation follows closely that of Nottale, sketched
into the previous section, but with some crucial
differences in the evaluation of the correlation functions
and the definition of the complex-time derivative operator,
respectively.

Before the derivation further comments on complex-energies
are in order. The energy functional $E_{QM}$ contains
imaginary components. Since meaningful physical solutions
demand real-valued energies this imposes constraints on the
physically acceptable states in these non-linear QM
equations, see Puszkarz \cite{starusz}.

Complex energy is not alien in ordinary linear QM. They
appear in optical potentials (complex) usually invoked to
model the absorption in scattering processes
\cite{starusz} and decay of unstable particles. Complex
potentials have also been used to describe decoherence
\cite{bars}. The accepted way to describe resonant states
in atomic and molecular physics is based on the complex
scaling approach, which in a natural way deals with complex
energies \cite{yanj}. We will show that real-valued energy
solutions exist to the NLSE based on a fractal Brownian
motion.

The imaginary part of the linear Schr\"odinger equation
yields the continuity equation $\partial\rho/\partial t +
\vec{\nabla}\cdot(\rho\vec{V}) = 0$. Because, as we shall
see, our potential is complex, the imaginary part of such
potential acts as a source term in the continuity equation.

Before, Nottale wrote,
\begin{equation}
\langle d\zeta_{\pm}d\zeta_{\pm}\rangle = \pm 2D dt,
\end{equation}
with $D$ and $2mD=\hbar$ real.

Now we set
\begin{equation}
\langle d\zeta_{\pm}d\zeta_{\pm}\rangle = \pm (D+D^*) dt,
\end{equation}
with $D$ and $2mD=\hbar=\alpha+i\beta$ complex.

The complex-time derivative operator becomes now
\begin{equation}
\frac{\delta}{dt} = \frac{\partial}{\partial
t}+\vec{V}\cdot\vec{\nabla} - \frac{i}{2}(D+D^*)\nabla^2.
\label{eq:complder}
\end{equation}
In the real case $D=D^*$. It reduces to the
complex-time-derivative operator described previously by
Nottale.

Writing again the $\psi$ in terms of the complex action $S$,
\begin{equation}
\psi=e^{iS/(2mD)}=e^{iS/\hbar},
\end{equation}
where $S$, $D$ and $\hbar$ are complex-valued, the complex
velocity is obtained from the complex momentum
$\vec{p}=\vec{\nabla}S$ as
\begin{equation}
\vec{V}=-2iD\vec{\nabla}\ln\psi.
\end{equation}

The NLSE is obtained after we use the generalized Newton's
equation (force = mass times acceleration) in terms of the
$\psi$ variable,
\begin{equation}
-\vec{\nabla}U = m\frac{\delta}{dt}\vec{V} =
-2imD\frac{\delta}{dt}\vec{\nabla}\ln\psi.
\end{equation}

Replacing the complex-time derivation (\ref{eq:complder})
in the generalized Newton's equation gives us
\begin{equation}
\vec{\nabla}U = 2im\left[D\frac{\partial}{\partial t}
\vec{\nabla}\ln\psi -
2iD^2(\vec{\nabla}\ln\psi\cdot\vec{\nabla})(\vec{\nabla}\ln\psi)
-\frac{i}{2}(D+D^*)D\nabla^2(\vec{\nabla}\ln\psi)\right].
\end{equation}
Now, using the three identities (1) $\vec{\nabla}\nabla^2 =
\nabla^2\vec{\nabla}$,
(2) $2(\vec{\nabla}\ln\psi\cdot\vec{\nabla})
(\vec{\nabla}\ln\psi) =
\vec{\nabla}(\vec{\nabla}\ln\psi)^2$ and (3) $\nabla^2\ln\psi
=\nabla^2\psi/\psi-(\vec{\nabla}\ln\psi)^2$ allows us to
integrate such equation above yielding, after some
straightforward algebra, the new NLSE that has the
nonlinear (kinematic pressure) potential found before
(\ref{eq:newNLSE}),
\begin{equation}
i\hbar\frac{\partial\psi}{\partial
t} = -\frac{\hbar^2}{2m}\frac{\alpha}{\hbar}\nabla^2\psi +
U\psi - i\frac{\hbar^2}{2m}\frac{\beta}{\hbar}
\left(\vec{\nabla}\ln\psi\right)^2\psi.
\label{eq:fNLSE}
\end{equation}
Note the crucial minus sign in front of the
kinematic pressure term and that $\hbar=\alpha+i\beta=2mD$
is complex. When $\beta=0$ we recover the linear
Schr\"odinger equation.

The nonlinear potential is now complex-valued in general.
Defining
\begin{equation}
W=-\frac{\hbar^2}{2m}\frac{\beta}{\hbar}
\left(\vec{\nabla}\ln\psi\right)^2,
\end{equation}
and $U$ the ordinary potential, then the NLSE can be
rewritten as
\begin{equation}
i\hbar\frac{\partial\psi}{\partial
t} =\left(-\frac{\hbar^2}{2m}\frac{\alpha}{\hbar}\nabla^2 +
U + iW\right)\psi.
\label{eq:ourNLSE}
\end{equation}
This is the fundamental nonlinear wave equation of this work.
It has the form of the ordinary Schr\"odinger equation with
the complex potential $U+iW$ and the complex $\hbar$. The
Hamiltonian is no longer Hermitian and the potential itself
depends on $\psi$. Nevertheless one could have meaningful
physical solutions with real valued energies and momenta.
Like the plane-wave and soliton solutions. Notice that the
new NLSE obeys the homogeneity condition $\psi\to\lambda\psi$
for any constant $\lambda$. All the terms in the NLSE are
scaled respectively by a factor $\lambda$. We did not
obtain the hydrostatic pressure term
$-b(\ln\psi^*\psi)\psi$ which breaks the homogeneity
condition for a simple reason: We are studying the true
kinematics and dynamics of a particle of mass $m$
undergoing a fractal Brownian motion. It would be
meaningless to have a hydrostatic pressure term in such a
model. Moreover, our two parameters $\alpha$, $\beta$ are
intrinsically connected to a complex Planck constant
$\hbar=\alpha+i\beta$ rather that being {\it ah-hoc\/}
constants to be determined experimentally. Thus, the
nonlinear QM equation derived from the fractal Brownian
motion with complex-valued diffusion coefficient is
intrinsically tied up with a non-Hermitian Hamiltonian and
with complex-valued energy spectra \cite{petrosky}.
To be more precise, the nonlinear $\beta$ term in
(\ref{eq:ourNLSE}) is really the
nonlinear partner of the kinetic energy term.

We will show that despite having a non-Hermitian
Hamiltonian we still could have eigenfunctions with real
valued energies and momenta. When $\hbar$ is real
($\beta=0$) and the NLSE is linearized back to the ordinary
one.

The reader may ask why not simply propose as a valid NLSE
the following,
\begin{equation}
i\hbar\frac{\partial\psi}{\partial
t} = -\frac{\hbar^2}{2m}\nabla^2\psi +
U\psi + \frac{\hbar^2}{2m}\frac{a}{m}
\vert\vec{\nabla}\ln\psi\vert^2\psi.
\label{eq:kpNLSE}
\end{equation}
Such equation (with the ordinary real Planck constant) is
based on a real Hamiltonian that satisfies the homogeneity
condition. It also admits soliton solutions of the type
\begin{equation}
\psi=C A(x-Vt)e^{i(kx-\omega t)},
\end{equation}
$A(x-Vt)$ is a function to be determined by solving the
NLSE. Galilean invariance imposes that the soliton is a
traveling wave, a function of $x-Vt$. We will present
explicit expressions for the function $A(x-Vt)$ afterwards.
Therefore, in principle, this NLSE is a more suitable
candidate that the Bia{\l}ynicki-Birula and Mycielski NLSE
with a nonlinear potential term $-b\ln(\psi^*\psi)$ that
breaks homogeneity and introduces also a $\psi^*$
dependence into the wave equation.

The only problem (perhaps there are others) with the NLSE
above is that it suffers also from an extraneous dispersion
relation. Plugging-in the plane-wave solution $\psi\sim
e^{-i(Et-px)/\hbar}$ one gets an extraneous energy-momentum
relation, after setting $U=0$,
\begin{equation}
E=\frac{\vec{p}^2}{2m}\left(1 +\frac{a}{m}\right),
\end{equation}
not the usual $E=\vec{p}^2/(2m)$. So in this case we have
that $E_{QM}\ne E_{FT}$ (FT means field theory).

It has been known for some time, see Puskarz
\cite{starusz}, that the expression for the energy
functional in nonlinear QM does not coincide with the QM
energy functional, nor it is unique. The simplest way to
see this is, for example, writing down the Birula and
Mycielski NLSE (\ref{eq:birula}) in the Weinberg form
\cite{weinberg}
\begin{equation}
i\hbar\frac{\partial\psi}{\partial
t} = \frac{\partial H(\psi,\psi^*)}{\partial\psi^*},
\label{eq:weinberg}
\end{equation}
where $\psi$ and $\psi^*$ are a pair of
canonically-conjugate variables. The real-valued
Hamiltonian density is given by
\begin{equation}
H(\psi,\psi^*) = -\frac{\hbar^2}{2m}\psi^*\nabla^2\psi +
U\psi^*\psi - b\psi^*\ln(\psi^*\psi)\psi + b\psi^*\psi,
\label{eq:hamdens}
\end{equation}
using $E_{FT}=\int d^3\vec{r}H$, so we can see it is
different from $\langle\hat{H}\rangle_{QM}$. Notice the
last term in (\ref{eq:hamdens}). Hence, one can immediately
see that the $H(\psi,\psi^*)$ stemming from the field
theory approach does not coincide with the Birula-Mycielski
Hamiltonian.  They differ by a constant, $E_{FT}-E_{QM}$ =
$\int d^3\vec{r} b \psi^*\psi$ = $b$. Exactly like it
occurs when we plug in the plane wave solution into the
NLSE with the nonlinear potential, real valued,
\begin{equation}
\frac{\hbar^2}{2m}\frac{a}{m}
\vert\vec{\nabla}\ln\psi\vert^2.
\end{equation}
Notice that this problem does not occur in the
fractal-based NLSE, because such NLSE is written entirely
in terms of the $\psi$ variable and does not contain the
$\psi^*$ variable explicit or implicitly, like it occurs in
the Birula-Mycielski NLSE.

The classic Gross-Pitaveskii NLSE (of the 1960'), based on
a quartic interaction potential energy, relevant to
Bose-Einstein condensation, contains the nonlinear cubic
terms in the Schr\"odinger equation, after differentiation,
$(\psi^*\psi)\psi$. This equation does not satisfy the
Weinberg homogeneity condition and also the $E_{FT}$
differs from the $E_{QM}$ by factors of two.

In the fractal-based NLSE there is no discrepancy between
the quantum-mechanical energy functional and the field
theory energy functional. Both are given by
\begin{equation}
H^{NLSE}_{fractal}=-\frac{\hbar^2}{2m}\frac{\alpha}{\hbar}
\psi^*\nabla^2\psi + U \psi^*\psi
-i\frac{\hbar^2}{2m}\frac{\beta}{\hbar}\psi^*
(\vec{\nabla}\ln\psi)^2\psi.
\end{equation}
The NLSE is then unambiguously given by equation
(\ref{eq:weinberg}), $H(\psi,\psi^*)$ is homogeneous of
degree 1 in $\lambda$ respect to $\psi$. This is why we
push forward the NLSE derived from the fractal Brownian
motion with a complex-valued diffusion coefficient. Such
equation does admit plane-wave solutions with the
dispersion relation $E=\vec{p}^2/(2m)$. It is not hard to
see that after inserting the plane wave solution into the
fractal-based NLSE we get (after setting $U=0$),
\begin{equation}
E=\frac{\hbar^2}{2m}\frac{\alpha}{\hbar}
\frac{\vec{p}^2}{\hbar^2} +
i \frac{\beta}{\hbar}\frac{\vec{p}^2}{2m} =
\frac{\vec{p}^2}{2m}\frac{\alpha+i\beta}{\hbar} =
\frac{\vec{p}^2}{2m},
\end{equation}
since $\hbar=\alpha+i\beta$. So the plane-wave is a
solution to the fractal-based NLSE (when $U=0$) with a
real-valued energy and which has the correct
energy-momentum dispersion relation.

\section{\bf Soliton solutions to the fractal based NLSE. One
dimensional case}
\label{sec:soliton}

Let us find soliton solutions to the fractal-based NLSE
given by (\ref{eq:fNLSE}), in the free particle case. We set
the ansatz (one-dimensional for simplicity)
\begin{equation}
\psi=C A(x-Vt)e^{-(Et-px)/\hbar}.
\end{equation}
The function $A$ must be complex-valued, otherwise no
real-valued energy solutions exist. Then we set
\begin{equation}
A(x-Vt)=F(x-Vt)+iG(x-Vt),
\end{equation}
and plugging-in $\psi$ with this $A$ into the fractal-based
NLSE (\ref{eq:fNLSE}) yields 2 coupled differential
equations, after separating the real and imaginary parts,
respectively, which yield, in principle, the functions $F$
and $G$.

For example, the soliton solution to the NLSE with the
$-b\ln(\psi^*\psi)$ is of the form \cite{pardy},
\begin{equation}
\psi(x,t)=C e^{a/B} e^{-(B/4)(x-Vt+d)^2} e^{ikx-i\omega t},
\end{equation}
where $c$, $a$, $B$, $d$ are numerical constants which can
depend on $\hbar$, $m$ and $b$.

As mentioned before, plane wave solutions to the NLSE based
on the $-b\ln(\psi^*\psi)$ potential exist but they have
extraneous dispersion relations. For example, the
energy-momentum relation turns out to be \cite{pardy} $E =
\hbar\omega = \vec{p}^2/(2m) + b\ln(2\pi)$. Thus,
plane-wave solutions do not seem to have physically
meaningful interpretation. This was another reason why we
believe that this NLSE has problems. Besides, we remarked
already that this NLSE breaks the homogeneity condition as
well.

To finalize we will find the soliton solutions to the NLSE
based on the kinematic pressure potential
$\vert\vec{\nabla}\ln\psi\vert^2$ terms, given by
equation (\ref{eq:kpNLSE}), in the free particle case
$U=0$.

Earlier on we have shown that it admits plane wave
solutions with the extraneous dispersion relation
$E=\vec{p}^2/(2m)(1+a/m)$. It obeys the homogeneity
condition: Under scaling of $\psi$ by $\lambda$ the NLSE
scales with an overall factor of $\lambda$ as expected.

Notice that if we wish to have a Hermitian Hamiltonian we
must take the absolute value
$\vert\vec{\nabla}\ln\psi\vert^2$ instead of
$(\vec{\nabla}\ln\psi)^2$ for our potential. Notice this
important difference between these non-linear potentials in
the fractal-based NLSE versus the kinematic pressure based
one.

Pluggin-in the ansatz $\psi=C F(x-Vt)e^{-(Et-px)/\hbar}$
into the kinematic pressure NLSE for the free particle case
$U=0$ yield for the imaginary parts $-i\hbar V F' =
-i(\hbar/m) F'p$. Then, for any $F$ we have $V = p/m$.
Therefore the ansatz is consistent with the de Broglie
relations $p=\hbar k$ and $E=\hbar\omega$, as expected from
this NLSE soliton solution.

The real parts give the differential equation
\begin{equation}
-F''F - \frac{a}{m} (F')^2 + \frac{1}{\hbar^2}\left[2mE-p^2
(1+\frac{a}{m})\right] F^2 = 0.
\end{equation}
The solutions to this nonlinear differential equation yield
$F(x-Vt)$. This differential equation involves the
derivatives $F'$, $F''$ and is much harder to solve than
the differential equation given in \cite{pardy} that
involves $F''$ but not $F'$.

For example a nonlinear differential equation which
involves $F''$ but not $F'$ is $F''-F^3=0$. Such equation,
after multiplying both sides by $F'$, can then be
integrated by quadratures, $\int dF/F^2 = \int dy/2^{1/2}$.

\section{\bf Concluding remarks}
\label{sec:conclu}

Based on Nottale and Ord's formulation of QM from first
principles; i.e. from the fractal Brownian motion of a
massive particle we have derived explicitly a nonlinear
Schr\"odinger equation. Despite the fact that the
Hamiltonian is not Hermitian real-valued energy solution
exist like the plane wave and soliton solutions in the free
particle case. The hydro-dynamical model analog of this
fractal-based NLSE yields another new NLSE with Hermitian
(real) Hamiltonian. The remarkable feature of the fractal
approach versus all the nonlinear QM equation considered so
far is that the quantum mechanical energy functional
coincides precisely with the field theory one.

The hydro-dynamical-based NLSE has a nonlinear (real)
potential term
\begin{equation}
\frac{a\hbar^2}{8m^2}
\vec{\nabla}\ln\frac{\psi}{\psi^*}\cdot\vec{\nabla}\ln\frac{\psi^*}{\psi},
\end{equation}
with $a$ the mass-energy parameter, bears a very rough
similarity to the Star\-usz\-kie\-wicz imaginary potential
term in three dimensions
\begin{equation}
-\frac{\gamma}{8}\frac{\nabla^2\ln\frac{\displaystyle\psi^*}{\displaystyle\psi}}{\psi^*\psi},
\end{equation}
see \cite{starusz}, this potential is imaginary, $\gamma$
is a constant.

The fractal model based NLSE admits plane wave (soliton
solutions also) with the correct dispersion relation
$E=\vec{p}^2/(2m)$, real. Soliton solutions, with
real-valued energy (momentum) are of the form
\begin{equation}
\psi\sim [F(x-Vt)+iG(x-Vt)]e^{ipx/\hbar-iEt/\hbar},
\end{equation}
with $F$, $G$ two functions of the argument $x-Vt$ obeying
a coupled set of two nonlinear differential equations.

It would be interesting to study solutions when one
turns-on an external potential $U\ne0$.

The reader may ask why concentrate on a complex diffusion
constant to generate a nonlinear Schr\"odinger equation
with a non-Hermitian Hamiltonian when one could have
written from the start a NLSE with a Hermitian Hamiltonian
that obeys the Weinberg homogeneity conditions and also
with the correct energy dispersion relations.

Starting from the fundamental equations: $\psi = e^{i
S/S_0} = e^{iS/\hbar_0}$ and the generalized Newtonian law,
written in terms of the Nottale complex derivative operator
and the $\psi$:
\begin{equation}
\vec{\nabla} U = i S_0 [ \frac{\partial\vec{\nabla}\ln\psi}{\partial t}
-i \{\frac{S_0}{m} (\vec{\nabla}\ln\psi\cdot\vec{\nabla})(\vec{\nabla}
\ln\psi) + D \nabla^2 (\vec{\nabla}\ln\psi) \} ],
\label{eq:newton}
\end{equation}
after adding and subtracting the quantity $D_0\nabla^2
(\vec{\nabla}\ln\psi)$, and using the 3 vector-calculus
identities used in (\ref{eq:ourNLSE}), we get a nonlinear
correction to the Schr\"odinger equation
\begin{equation}
\frac{\hbar_0}{2m}(\hbar - \hbar_0) (\nabla^2\ln\psi) \psi,
\end{equation}
where $S_0 = \hbar_0 = 2m D_0$ and $\hbar = 2mD \not=
2mD_0$.

As desirable as this NLSE may look a close inspection
reveals that the nonlinearity is just an artifact of the
definition of $ \psi $. It looks nonlinear from the $\psi$
perspective. It is not difficult to see that under a
re-definition of the wavefunction
\begin{equation}
\psi' = e^{i S/\hbar} = e^{i S/(2mD)}
\end{equation}
the generalized Newtonian law of motion of a particle
undergoing a fractal Brownian motion given by
(\ref{eq:newton}) yields now the standard linear
Schr\"odinger equation for the $ \psi'$ wavefunction. Where
we have chosen for a new value of $ S_0' = \hbar = 2mD$.

It is important to emphasize that the diffusion constant is
always chosen to be related to Planck constant as follows:
$2 m D = \hbar$ which is just the transition length from a
fractal to a scale-independence non-fractal regime
discussed by Nottale in numerous occasions. In the
relativistic scale it is the Compton wavelength of the
particle (say an electron): $\lambda_c = \hbar/(mc)$. In
the nonrelativistic case it is the de Broglie wavelength of
the electron.

Therefore, the NLSE based on a fractal Brownian motion with
a complex valued diffusion constant $2mD = \hbar = \alpha +
i \beta $ represents truly a new physical phenomenon in so
far as the small imaginary correction to the Planck
constant (unobserved in present day experiments) is the
hallmark of nonlinearity in QM. For other generalizations
of QM see experimental tests of quaternionic QM (in the
book by Adler \cite{adler}).
Equation (\ref{eq:ourNLSE}) is the fundamental NLSE of this work, where
the $\beta$ term is essentially the nonlinear partner of the linear
kinetic energy term in comparison to all other approaches which focused on
nonlinear modifications of the potential.

\section*{\bf Acknowledgements}

We acknowledge to the Center for Theoretical Studies of
Physical Systems, Clark Atlanta University, Atlanta,
Georgia, USA, and the Research Committee of the University
of Antioquia (CODI), Medell\'{\i}n, Colombia for support.
C. C. wishes to thank Dr. Alfred Sch\"oller and family for
their Austrian hospitality.

\end{document}